\documentclass[12pt]{article}
\usepackage{graphicx,amsmath}
\usepackage{units}

\def\be{\begin{equation}}
\def\ee{\end{equation}}
\begin{document}

\vspace*{0.5cm}
\begin{center}
{\Large \bf The diffraction cone shrinkage  speed up with the collision energy  }\\

\vspace*{1cm}

V.A. Schegelsky, M.G. Ryskin \footnote{e-mail addresses: valery.schegelsky@cern.ch,misha.ryskin@durham.ac.uk}\\

\vspace*{0.5cm}
 Petersburg Nuclear Physics Institute, Gatchina, St.~Petersburg, 188300, Russia \\

\vspace*{1cm}

\begin{abstract}

The multiperipheral ladder structure of the Pomeron leads to the  quite natural conclusion that the elastic slope
$B_{el}$ is not simple linear function of the colliding  particles energy  logarithm. The existing experimental data
on the diffraction cone shrinkage points to such "complicated" dependence indeed. The  diffraction
 cone shrinkage speed up with the beam energy is
directly connected with an extreme rise of total cross-section ( Froissart limit).

\end{abstract}
\vspace*{0.5cm}

\end{center}

\section{Introduction}
 At high energies the hadron-hadron scattering are usually described by the Pomeron exchange. A popular
 parameterization of the elastic scattering amplitude at small momentum transfer takes into account only
  Reggeon and
Pomeron poles exchange . The $ab$ elastic scattering amplitude reads
\be
T_{ab}(t)=F_a(t)F_b(t)C_Ps^{\alpha_P(t)}+F_R(t)C_Rs^{\alpha_R(t)}
\ee
where the form factors $F_a$, $F_b$, $F_R$ describes the matter distribution in the incoming hadrons $a,b$. $C_P$ and
$C_R$ are the normalization constants. The contribution
of the secondary Reggeon poles (last term in (1)) becomes negligible at $\sqrt{s} \sim 100 GeV$. \\

 From the microscopic viewpoint the Pomeron is described by the
 ladder-type diagrams in which the energy
(longitudinal momentum fraction) in each next cell is a few times smaller  than that in the previous cell
\footnote{This multiperipheral ladder structure of the Reggeon was considered first in \cite{AFS}.}.
 To get
the largest cross section we have to consider the chain (sequence) of strong  interactions with relatively low
partial sub-energies. Such a sequence of interactions provides a large
 -non-decreasing with energy cross section $\sigma\propto  s^{\alpha_P(0)-1}$.\\

On another hand at each step the interaction radius changes by the value $\delta\rho\sim 1/k_t$ leading to the
'diffusion' in impact parameter plane. At each step the energy of incoming particle diminishes a few times. Thus the
number of steps is
 $n\sim \ln s$ and the final radius is
 $R^2=R^2_0+n\cdot(\delta\rho)^2$.

Therefore the Pomeron trajectory $\alpha_P(t)$ depends on the momentum transferred $t=-q^2_t$
and for a not large $|t|$ it can be written as $\alpha_P(t)=1+\epsilon+\alpha'_P t$.\\

Correspondingly the elastic $ab$-cross section takes the form
\be
\frac{d\sigma_{ab}}{dt}=\frac{\sigma_0^2}{16\pi}F_a^2(t)F_b^2(t)\left( \frac s{s_0}\right)^{2\epsilon+2\alpha'_P t}.
\label{sgma-el}
\ee

The power growth of the "single Pomeron exchange" cross section  generated by the ladder diagram
reflects the growth of the parton
 multiplicity, $N$.
Since at each (ladder) step the longitudinal momentum decreases by a few times the mean number of steps $<n>\sim
c*\ln s$. At each splitting (step) the multiplicity of parton increases by a factor two. Thus the final  parton
multiplicity $N\sim 2^{c\ln s}=s^{c\ln 2}$.

The slope of Pomeron trajectory $\alpha'_P$ accounts for the growth of interaction radius caused by a long chain of
intermediate (relatively low energy) interactions which length increases with $\ln s$. In the case of Gaussian form
factors $F_a^2F_b^2=\exp(B_0t)$ we get the slope of elastic cross section $d\sigma/dt=|T(t)|^2/16\pi s^2\propto
\exp(B_{el}t)$
\be
B_{el}=B_0+2\alpha^{' eff}_P\ln(s/s_0).
\label{b-el}
\ee
While the first term $B_0$  in (\ref{b-el}) depends on the sort of incoming hadrons $a$ and $b$,the second term
$2\alpha^{' eff}_P\ln(s/s_0)$ is universal. In the case of one Pomeron exchange it should be the same at any energy
and for any type of incoming hadrons. This universality was confirmed at the fixed target
 experiments\cite{Burq}( $\sqrt{s}$ = 24 GeV)  where the value of $\alpha'_P=0.14$ GeV$^{-2}$ was measured.
 \footnote{At a not too large fix target energies it was important to account for the secondary
  Reggeon contribution in fit \cite{Burq}.} \\

Donnachie-Landshoff \cite{DL}(equation 7) have deduced   from the shape
of $d\sigma_{el}/{dt}$ at measured  $\sqrt{s}$ = 52.8 GeV \cite{ISR}  much  larger $\alpha'_P=0.25$ GeV$^{-2}$.\\

\section{More complicated slope $B_{el}(s)$ behavior }
The  growth of $\alpha^{' eff}_P$ should be expected indeed. When the optical (parton) density, i.e. the opacity
$\Omega(\rho,s)$, becomes too large we have to account for the multiple interactions which are described by the
multi-Pomeron diagrams. Like in the case
 of nuclear-nuclear $AA$-collisions, where few nucleon-nucleon
pairs may interact simultaneously and screen each other, the corresponding absorptive
 corrections stop the growth of elastic amplitude near the black
 disk limit, when in impact parameter representation the
imaginary part of the elastic scatterng amplitude $Im T(\rho)\to 1$.
 \footnote{Recall that these
 multi-Pomeron diagrams are generated just by the $s$-channel two
 particle unitarity. Within the eikonal model, the
 amplitude given by the sum of multi-Pomeron contributions reads
 $T(\rho)=i(1-\exp(-\Omega(\rho,s)/2))$ where the value of $\Omega$
 is described by the one Pomeron (ladder) exchange.}.
Note that while at the center of the disk ( at small $\rho$) the amplitude saturates at $ImT=1$, it still continues to
increase with energy at the periphery (at large $\rho$) leading to the growth of the mean interaction radius and thus
to the growth of $t$-slope $B_{el}$. {Another way to see the variation of $B_{el}$ with energy is to consider just
two first diagrams - the one Pomeron exchange and the two Pomeron cut. In comparison with the one Pomeron exchange
the two Pomeron contribution falls down with $-t$  slowly, since using two (few) Pomerons we may distribute the whole
transferred momentum more homogeneously between the components (partons) of the initial hadron. However the two
Pomeron contribution describes the absorptive correction and has the  sign opposite to that of the one Pomeron
exchange. Therefore the $t$ dependence of the whole amplitude becomes steeper and the slope $B_{el}$ increases  for
the case of $\alpha_P(0)>1$, when at  larger energies the relative size of the two Pomeron cut increases.}

Therefore the effective shrinkage of the diffractive cone is described by the value of $\alpha^{' eff}_P$ which
accounts for both- the growth of the radius of individual Pomeron ( $\alpha'_P$ of the 'bare' Pomeron trajectory) and
the decrease of  optical density in the center of the disk (in comparison with the periphery) because of  absorptive
effects which provide the radius growth with energy .
So $\alpha^{' eff}_P>\alpha'_P$. \\

\section{ The increase  $\alpha^{' eff}_P$ with collisions energy}
\begin{figure}
\includegraphics[height=8cm]{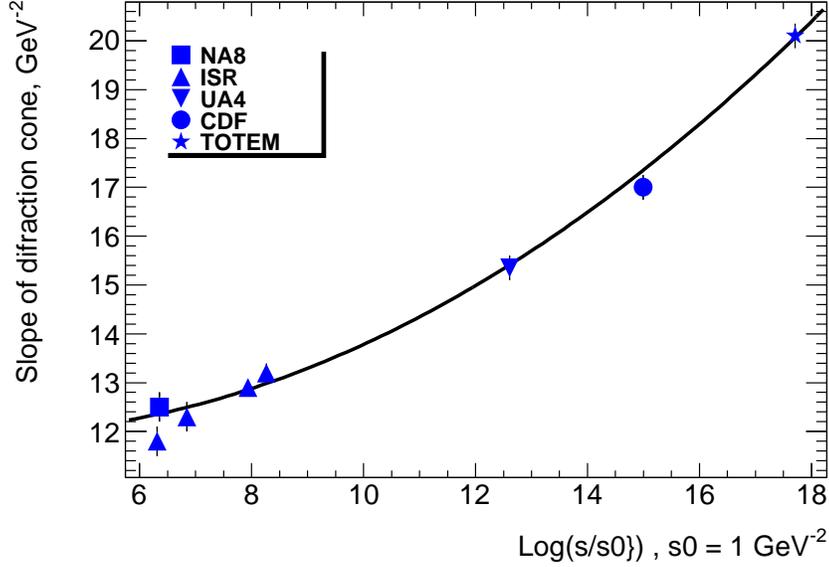}
\caption{\sf The existing measurement of the diffraction cone slope $B_el$. }
\label{fig:slope}
\end{figure}

Fig.~\ref{fig:slope} shows measured values  of the elastic $t$-slope $B_{el}(s)$  ( NA8-Gatchina-Cern \cite{Burq},ISR
\cite{ISR}, UA4 \cite{UA4}, CDF \cite{CDF})
 including new TOTEM result \cite{TOTEM}.

 One can see clearly that the value of $\alpha^{' eff}_P$ does grow with energy.
 Fitting the $\ln s$ dependence of $B_{el}$ by the second order
 polynomial
\be
B_{el}=B_0+b_1\ln(s/s_0)+b_2\ln^2(s/s_0)
\label{b4}
\ee
we get

$b_1=(-.22 \pm .17)GeV^{-2}$ and
 $b_2=(.037 \pm .006) GeV^{-2}$

 with good $\chi^2/NoF =7.5/5$ while fit with the linear function is unacceptable $\chi^2/NoF =37.8/6$.
 In all fits we use $s_0=1$ GeV$^2$. Recall
that the coefficient $b_2$ (and the analogous coefficient $c_2$ in the expression  for the total cross section in
sect.4) does not depend
 on the value of $s_0$. Changing $s_0$ we only re-define the coefficients $B_0$ and $b_1$ .
  Moreover at a given beam energy the value of
2$ \alpha^{' eff}_P = dB_{el}/d(ln(s/s_0)) $
 is also independent on the scale $s_0$.\\
 Note that in the case of $s_0=1$ GeV$^2$ the value of $b_1$   is consistent  with zero.
 The exclusion of this parameter and the fit with the function
\be
B_{el}=B_0+b_2\ln^2(s/s_0)
\label{b5}
\ee
gives
\be
 b_2=(0.02860 \pm 0.00050)  GeV^{-2}
\ee
 and does  not change statistical significance :  $\chi^2/NoF = 3.3/4$ against of
$\chi^2/NoF =3.9/5$.

\begin{figure}
\includegraphics[height=8cm]{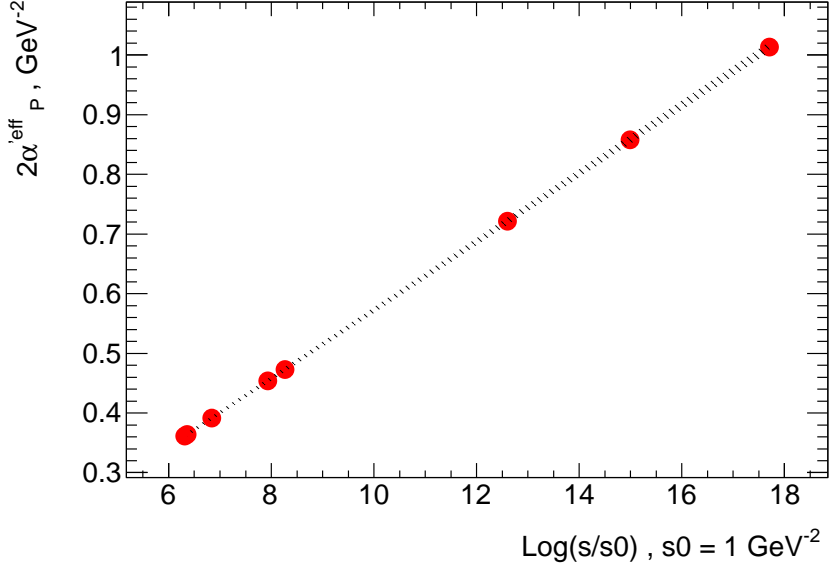}
\caption{\sf  The energy dependence of 2$\alpha^{' eff}_P$. }
\label{fig:alpha}
\end{figure}

The energy dependence of 2$\alpha^{' eff}_P = dB_{el}/d(ln(s/s_0)) $ is shown in Fig.~\ref{fig:alpha}

\section { Froissart limit for the diffraction cone shrinkage}
 Let us compare the behavior of the slope $B_{el}$ and the total $pp$ cross section in the asymptotic black
disk (Froissart) limit, when $\sigma_{tot}=2\pi R^2$ and $B_{el}=R^2/4$ (here $R$ is the black disk radius).

The recent fit $\sigma_{tot}=\sigma_0+c_1\ln(s/s_0)+c_2\ln^2(s/s_)$ gives $c_2=(0.2817 \pm 0.0064)$ mb (see Table 1
of \cite{Block})while from $b_2=0.037$ GeV$^2$ we get $c_2(B_{el})=0.375$ mb
 and from $b_2=(0.0286\pm 0.0005) GeV^2$, obtained in two parameters fit,
 we get a very close value -- $c_2(B_{el})=(0.294\pm 0.005)$ mb.
 This demonstrates the present uncertainty in the coefficent $c_2$
 extracted from the elastic slope behaviour.
Of course, even at the LHC we are rather far from the complete black disk limit. The proton is still relatively
transparent and the cross section $\sigma_{tot}$ is less than its geometric value $2\pi R^2$.

 However it is interesting that both the elastic $t$-slope and the total cross section have the
same $\ln^2 s$ high energy behavior. Starting from the elastic slope we get from the coefficient $b_2$ the value of
$c_2$ close to that obtained from the total cross sections.

 Nontrivial fact is that the value of 2$\alpha^{' eff}_P = (0.26\pm0.17) GeV^{-2}$ for 3-parameters fit or
2$\alpha^{' eff}_P = (0.364\pm0.003) GeV^{-2}$  for two parameters fit at $\sqrt{s} = 24$ GeV are similar to
$2\alpha'_P = (0.28\pm0.03) GeV^{-2}$ found in the Regge Poles analysis of " low energy" elastic scattering
\cite{Burq}.

   Unfortunately, our conclusion about the non-linear $\ln(s)$
 behaviour of the slope $B_{el}$ is based (besides the  Regge
 {\em Theory}) on the only ONE {\em measurement} - TOTEM \cite{TOTEM}.
   It looks that in the energy region $\sqrt{s} = 2-7$ TeV the
role of multi-Pomeron contributions strongly increases.
The multi-Pomeron effects should reveal itself not only in elastic scattering but
in the multiparticle production as well (see the discussion in \cite{KMR}).

Recall that the recent Donnachei-Landshoff fit\cite{DL2} includes {\em two} Pomeron poles. The pole with high intercept $\epsilon=0.362$ and the pole with $\epsilon=0.093$. Each of these 'effective' poles should produce its own secondaries and it would be important to observe the two different power of $s$ in the behaviour of the inclusive cross sections, $d\sigma/d^3p$, and in two particle correlations, including the Bose-Einstein correlations where these two poles will act as two different sources of secondary mesons. Since the slope of the trajectory with a higher intercept is smaller than that for the pole with $\epsilon=0.093$, we expect that
the emission size corresponding to the pole with $\epsilon=0.362$
should be smaller as well.

 Only the LHC can investigate this energy region performing
the energy scan in the manner previously
 realized with the S$p\bar p$S collider.


\thebibliography{}
 \bibitem{AFS} D. Amati, S. Fubini, A. Stanghellini,
 Nuovo Cim. {\bf 26}, 896, (1962).
\bibitem{Burq} J.P.Burq et al ,Nuclear Physics {\bf B217} (1983) 285-335.
\bibitem{DL}  A. Donnachie, P.V. Landshoff,  Nucl.Phys. {\bf B231} (1984) 189.
\bibitem{ISR} N.Amos et al, Nuclear  Physics {\bf B262} (1985) 689-714.
\bibitem{UA4} C. Augier, et al.,Phys. Lett. {\bf B316} (1993) 448.
\bibitem{CDF}Phys Rev, {\bf D50} (1994) 5518.
\bibitem{TOTEM} TOTEM Coll , Eur.phys. Lett. {\bf 96} (2011) 21002.
\bibitem{Block}   
  Martin M. Block,  Francis Halzen,
e-Print: arXiv:1109.2041 [hep-ph]
\bibitem{KMR}
  M.G. Ryskin, A.D. Martin, V.A. Khoze, A.G. Shuvaev,
 J.Phys.G {\bf 36} (2009) 093001; M.G. Ryskin, A.D. Martin,
V.A. Khoze, J.Phys.G {\bf 38} (2011) 085006
\bibitem{DL2}  A. Donnachie, P.V. Landshoff, arXiv:1112.2485 [hep-ph]
\end{document}